\documentclass[conference]{IEEEtran}
\IEEEoverridecommandlockouts
\usepackage{cite}
\usepackage{amsmath,amssymb,amsfonts}
\usepackage{algorithmic}
\usepackage{graphicx}
\usepackage{textcomp}
\usepackage{xcolor}

\usepackage[caption=false,font=footnotesize,labelfont=sf,textfont=sf]{subfig}

\begin{document}

\title{Peer-to-Peer EnergyTrade: A Distributed Private Energy Trading Platform \\
\thanks{The first and second authors have the same contribution in the paper.}
}

\author{\IEEEauthorblockN{ Ali Dorri}
\IEEEauthorblockA{\textit{CSE, UNSW} \\
Sydney, Australia \\
ali.dorri@unsw.edu.au}
\and
\IEEEauthorblockN{Ambrose Hill}
\IEEEauthorblockA{\textit{CSE, UNSW} \\
Sydney, Australia \\
ambrosewhill@gmail.com}
\and
\IEEEauthorblockN{ Salil S Kanhere}
\IEEEauthorblockA{\textit{CSE UNSW} \\
Sydney, Australia \\
salil.kanhere@unsw.edu.au}\\
\and
\IEEEauthorblockN{Raja Jurdak}
\IEEEauthorblockA{\textit{DATA61 CSIRO} \\
	Brisbane, Australia. \\
	Raja.Jurdak@csiro.au}\\
\and 
\IEEEauthorblockN{Fengji Luo}
\IEEEauthorblockA{\textit{University of Sydney} \\
	Sydney, Australia. \\
	fengji.luo@sydney.edu.au }\\
\and 
\IEEEauthorblockN{ Zhao Yang Dong }
\IEEEauthorblockA{\textit{UNSW} \\
	Sydney, Australia. \\
	joe.dong@unsw.edu.au }
}

\maketitle

\begin{abstract}
Blockchain is increasingly being used as a distributed, anonymous, trustless framework for energy trading in smart grids. However, most of the existing solutions suffer from reliance on Trusted Third Parties (TTP), lack of privacy, and traffic and processing overheads. In our previous work, we have proposed a Secure Private Blockchain-based framework (SPB) for energy trading to address the aforementioned challenges. In this paper, we present a proof-on-concept implementation of SPB on the Ethereum private network to demonstrates SPB's applicability for energy trading. We benchmark SPB's performance against the relevant state-of-the-art.     The implementation results demonstrate that SPB incurs lower overheads and monetary cost for end users to trade energy compared to existing solutions.
\end{abstract}


\section{Introduction}
Power systems are experiencing profound changes  with the  penetration of distributed renewable energy sources and energy storage systems, deployment of advanced metering and sensing facilities, and participation of flexible power loads.  With these technical developments, traditional energy consumers in power distribution networks (e.g.,  residential, industrial, commercial buildings) are increasingly being transformed to energy \textit{prosumers} (producers-and-consumers), meaning that they are capable of both generating and consuming energy. As a result, in recent years there has been interest from both academia    \cite{long2017peer,wang2015incentive,luo2018distributed} and industry \cite{powerledger,Conjoule,LO3Energy} on facilitating peer-to-peer energy trading among prosumers in  power distribution. Alongside its potential benefits, peer-to-peer energy trading involves challenges of security, privacy, and reliance on Trusted Third Parties (TTPs). \par 
 
Blockchain has significant potential to underpin a distribution energy trading solution  due to its salient features  including decentralization, security, auditability, and anonymity. In blockchain, interactions  between nodes are known as transactions. Particular nodes in the network, known as miners, periodically collect pending transactions and form  a new block by following a consensus algorithm. The latter ensures blockchain security against malicious miners to achieve distributed trust in trustless network. Proof of Work (POW) \cite{nakamoto2008bitcoin} and Proof of Stake (POS) \cite{king2012ppcoin} are examples of such algorithms. The miners receive incentive for storing transactions in blockchain in the form of \textit{transaction fee } paid by the transaction generators. All transactions and blocks are broadcast to and verified by all participants which  eliminates the need for central controllers and achieves decentralization.  All  transactions are cryptographically sealed using asymmetric encryption. The Public Key (PK) associated with each transaction serves as the identity of the transaction generator. Each user can change his PK per transaction to  enhance his anonymity level. \par 

The authors in \cite{mengelkamp2018designing}  demonstrated the applicability of  blockchain for energy trading by presenting a proof-of-concept implementation.  In \cite{laszka2017providing} the authors proposed a blockchain-based energy trading platform where  energy is converted to assets and the asset can be traded in the blockchain. Beyond the research community, blockchain has also received significant attention from industry.  Powerledger  \cite{powerledger} proposed a blockchain-based energy market place.  However, the existing blockchain-based solutions for energy trading suffer from the following challenges: i) Reliance on Trusted Third Party (TTP) brokers to ensure that both sides of energy trade fulfill their commitments, ii) Lack of privacy as attackers can obtain critical privacy-sensitive information of the user by linking multiple transactions or monitoring the transaction generation patterns of nodes, and iii) Blockchain overheads as negotiations between energy producers and consumers are broadcast to all participants.\par 

We recently proposed   a Secure Private Blockchain-based (SPB) energy trading framework to address the aforementioned challenges \cite{dorri2018spb}. To eliminate TTP, SPB introduces atomic meta-transactions where a transaction is considered to be valid if and only if it is coupled with another transaction. The atomic meta-transactions are: 1) Commit To Pay (CTP): generated by the consumer to commit to pay the energy price to the producer, 2) Energy Receipt Confirmation (ERC): generated by the smart meter of the consumer to confirm receipt of the energy. To verify ERC, other participants in the blockchain need to verify that the ERC is generated by a genuine smart meter which protects against malicious nodes that may generate fake ERC transactions by claiming to be a smart meter.  However, the transactions generated by the smart meter may reveal privacy-sensitive information of the smart meter owner, e.g., energy consumption pattern \cite{varodayan2011smart}. To address this challenge, SPB proposed the notion of a  Certificate of Existence (CoE). The CoE is the root hash of a Merkle tree that is signed by another meter in the network (see Section \ref{subsec:spb} for further details). Each smart meter constructs the Merkle tree  by recursively hashing  a particular number of PKs. The smart meter populates the CoE in  ERC transactions and uses one of the PKs from the Merkle tree and its  corresponding signature. The latter proves that the ERC generator is the original generator of the Merkle tree and thus is a smart meter.   In the existing negotiation approaches between energy producer and consumer all messages associated with negotiation of energy trade   are manifested as transactions and thus are broadcast to all nodes in the blockchain.  To reduce the associated overhead during negotiation, SPB introduces a new routing algorithm on top of blockchain which unicasts negotiation transactions  between energy producer and consumer.\par 

The main contribution of the current paper is to  evaluate a Proof-of-Concept  (PoC) implementation of SPB. We implement the core SPB functions on top of the Ethereum private network. We use two Raspberry Pi3s  where one mimics the smart meter of the energy consumer and the other acts as the solar panel of the energy producer.  We utilize a standard Macbook Pro 2015 as the miner. The implementation shows the applicability of SPB for distributed energy trading. The results demonstrate SPB reduces  delay and monetary cost  as compared to the existing solutions from the end user perspective.  We also show that SPB reduces the blockchain memory footprint by 40\%, and the associated end-to-end delay in energy trading by up to 35\% as compared to a baseline method. The results show that SPB reduces  the associated network traffic and processing overheads, thus demonstrating its scalability.  \par 

The rest of the paper is organized as follows. Section \ref{sec:background} discusses the necessary background which includes a literature review of energy trading in  Section \ref{subsec:relatedwork}, and an overview of SPB in Section \ref{subsec:spb}. Section \ref{sec:implementation setup} outlines the implementation setup and discusses the  results. Finally, Section \ref{sec:conclusion} concludes the paper and outlines future works.

\section{Background}\label{sec:background}
In this section, we  discuss the related work in energy trading, followed by a detailed discussion of SPB. 

\subsection{Distributed Energy Trading} \label{subsec:relatedwork}

The authors in \cite{aitzhan2018security}  proposed a blockchain-based energy trading platform where the energy producer and consumer can negotiate the energy price and action the energy trade.  The proposed method is built on top of Bitcoin and uses Proof of Work (POW) as the underlying consensus algorithm. The energy producer and consumer may negotiate the price and amount of energy by broadcasting negotiation transactions which are  encrypted by the PK of the destination to  enhance the  privacy of the energy consumer and producer.   The energy is transfered to an asset and the ownership of the asset is traded in the blockchain. To protect against malicious energy producers that may attempt to sell a given block of energy  to multiple consumers (essentially a form of double spending \cite{karame2012double}),  when a producer sells the asset, i.e., energy, it must lock the asset by  sending a message to the  energy company. This message contains the ID of the asset and is signed by the producer. On receiving the message, the energy company flags the asset as sold.  The consumer verifies with the energy company to ensure that the asset is flagged as sold. Other consumers can verify if the asset is sold by querying the energy company prior to energy trading.  Multisign transactions are used, wherein,  a transaction must at least have 2 out of 3 signatures to be considered as a valid transaction. The signatories must be the producer, consumer, and the energy company. The latter is added to arbitrate any disputes that may eventuate between the consumer and producer.\par

The authors in \cite{mihaylov2014nrgcoin} proposed to convert energy into a currency, named NRGcoin, which is then transferred in the blockchain. According to  demand and load in the network, the price of the NRGcoin, i.e., the price of the energy, is determined by  Distributed System Operator (DSO). The latter is essentially the energy company. The participating nodes in energy trading submit their load and demand requests to their local  street-level  energy substation of DSO which are  used to determine NRGcoin price.  \par 

In \cite{li2018consortium} the authors proposed \textit{energy blockchain} as a unified blockchain-based platform for Industrial Internet of Things (IIoT) which supports  multiple energy trading scenarios, e.g., vehicle-to-grid and smart homes. To reduce the delay associated with transaction confirmation, a credit-based payment method is proposed, where a central trusted bank manages the credits and payments during energy trading. Similar to conventional banks, the proposed bank tracks credit transfer from its users' account thus  eliminates the requirement for transaction confirmation. The participating nodes can also borrow credit from the bank based on their credit history.\par 

The authors in \cite{laszka2017providing} proposed a framework to trade energy in a distributed manner using blockchain. The energy bids and requests are stored in a central database that is then used by the producers and consumers to find a match to sell/buy energy. To protect privacy,  the PK of the producer or consumer  is sent to a mixing service which assigns a new completely random PK to the prosumer. \par 

The authors in \cite{DistributedAPP} proposed a distributed solution for trading goods including energy using Ethereum that relies on smart contracts. Once the  buyer and seller agreed on the price, the buyer pays the price of the good to the smart contract. Once the payment transaction is stored in the blockchain, the seller transfers the goods  to the buyer. The buyer confirms receipt  by sending a confirmation to the smart contract which triggers the smart contract to pay the price to the seller.  \par 

Distributed energy trading has also received attention from industry.  Solara \cite{Solara} proposed a blockchain platform where participants can verify whether energy is generated by renewable energy resources. A Solara Hardware Module (SHM) is  utilized to confirm the energy generation source.   Powerledger  \cite{powerledger} combines permissionless, i.e., public, and permissioned blockchain, i.e., private where only authorized nodes can join the blockchain.  The energy prosumer buys tokens in a public Ethereum-based blockchain  which authorizes it to join a permissioned blockchain for energy trading. \par 
Collectively, the state-of-the-art blockchain-based energy trading solutions (including those outlined above),  suffer from the following limitations:
\begin{itemize}
	\item \textit{Lack of privacy:} Blockchain achieves some level of anonymity as the users employ changeable   PKs as their identity. However, transactions with different PKs can be linked together to deanonymize a user and thus compromise his privacy \cite{dupont2015toward}. Recent studies in Bitcoin shows that using  changeable PKs or mixing services does not secure the users against deanonymization and the same concept can be applied in energy trading \cite{hong2018practical}. Further, in energy trading,  monitoring the pattern of transaction generation may reveal privacy-sensitive information about the user such as energy consumption and production patterns \cite{varodayan2011smart}.  	In most of the existing solutions, transactions generated by a user can be tracked back to the user (e.g., \cite{aitzhan2018security,mihaylov2014nrgcoin}).  
	\item \textit{Reliance on TTP: } In energy trading,  both sides of the trade must fulfill their  commitments, i.e., the producer must send the committed amount of energy to the consumer and the latter must in turn pay out the agreed price to the former. Achieving this trust  in blockchain is challenging due to its distributed nature. To address this challenge,  most of the existing works rely on TTPs that carefully monitor the entire energy trading process. The TTP can be the energy company \cite{mihaylov2014nrgcoin} or a central bank \cite{li2018consortium}.  However, TTP suffers from the inherent limitations associated with centralization. The TTP also may compromise the privacy of the participants as it monitors the details of trades.  
	
	\item \textit{Blockchain overheads: } Some of the existing works such as \cite{aitzhan2018security} employ POW and Bitcoin as the underlying consensus and payment solutions. Solving the hard to solve cryptographic puzzles employed by these algorithms results in significant overheads  which in turn consumes substantial energy  \cite{o2014bitcoin}. Recent reports show that Bitcoin energy consumption equals with energy consumption of Ireland \cite{BitcoinEnergy}.  To negotiate the energy price, the existing methods  use broadcast messages with destination address (similar to Ethereum whisper \cite{Ethereumwhisper,aitzhan2018security}) that  incurs significant packet overhead on the blockchain and consumes significant resources.
\end{itemize}
To address these limitations in this paper we base our study on SPB which is designed with the goal of overcoming the aforementioned issues. We provide a brief overview of SPB  in the next section as background for the proposed implementation. Full details of SPB are available in \cite{dorri2018spb}.\par 
\subsection{Secure Private Blockchain-based energy trading: SPB}\label{subsec:spb}

SPB enables energy consumers and producers to negotiate and trade energy in a distributed manner while eliminating the need for a TTP. The participating nodes in the smart grid including energy producers, consumers, prosumers, and distribution companies jointly manage the blockchain by storing and verifying transactions and blocks.  The producer instantiates an \textit{energy account} which is a ledger of transactions generated by the producer. To generate the energy account, the user either has to burn coin in Bitcoin, i.e., pay specific amount to an unknown address, or receive certificate from authorities, e.g., energy companies. The producer progressively adds blocks of energy to his account, as they are generated,   along with the energy price. To do so, the producer deploys a smart contract that can be used for all its energy tradings. A producer may also utilize the smart contract generated by another producer  as smart contracts share the same functions. The smart contract maintains the amount and price of energy for energy producers. \par 

\begin{figure}
	\centerline{\includegraphics[width=9cm,height=9cm,keepaspectratio]{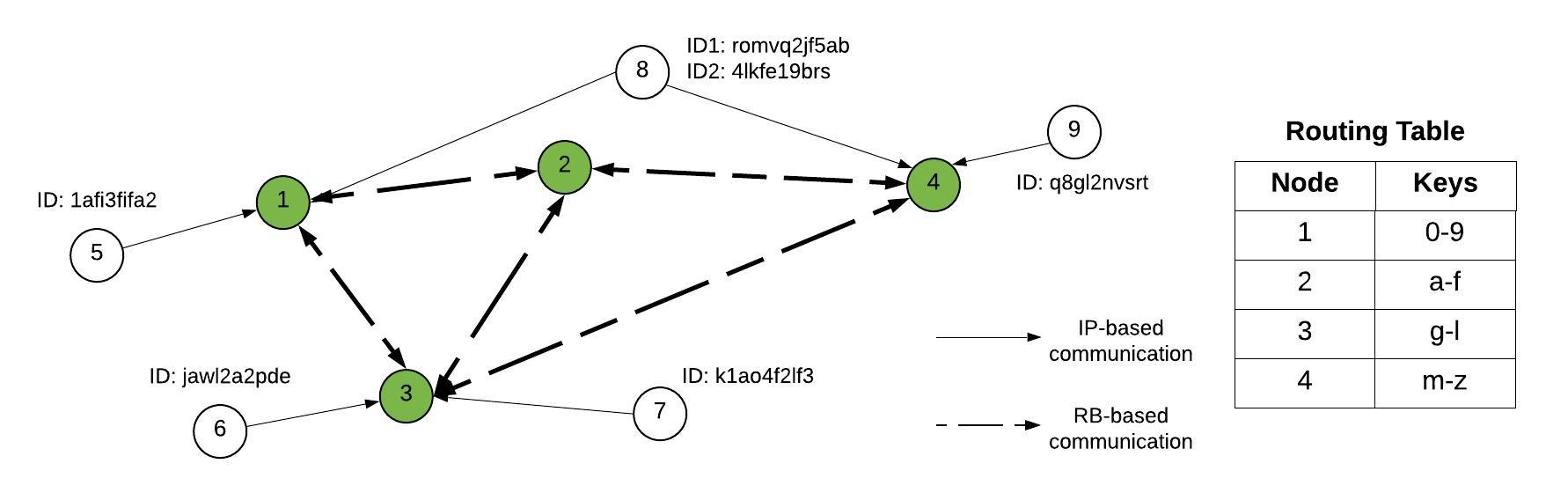}}
	\caption{An example of routing packets using backbone nodes  \cite{dorri2018spb}.}
	\label{fig:routing}
\end{figure}

The energy consumer initiates a query to search for  the available energy and price by exploring the blockchain. This search is similar to searching for unspent transactions in Bitcoin. The consumer can negotiate the price of energy with the producer if the price is negotiable. To do so, SPB proposes a new routing approach that uses PKs as identifiers.  The participating nodes that have high resources available form a backbone network that is responsible for routing negotiation transactions. The backbone nodes route packets based on the value of the most significant bites of the PK of the destination.  Figure  \ref{fig:routing} depicts an illustrative example of the entire process. The   participating nodes in the network  associate with the backbone node that is responsible for their PK. Recall that in blockchain a user might employ multiple PKs to protect his anonymity. Thus, a node might be associated with multiple backbone nodes for different keys, e.g., node 8 in Figure \ref{fig:routing}.\par 
After negotiation (if any) the consumer and producer resume the energy trade process. To eliminate the need for TTP, SPB introduces the notion of atomic meta-transactions which comprises the following two transactions that must be lodged in sequence within a certain time:
\begin{itemize}
	\item \textit{Commit To Pay (CTP): }This transaction is generated by the energy consumer to commit to pay the energy price to the producer. The CTP does not transfer money to the consumer account, but rather puts  the money on hold till the energy trading concludes. Conceptually, this is similar to putting money on hold on credit cards. If the producer does not transfer the energy to the consumer within a certain time period, then this money must be released. To achieve this an \textit{expiry time} is included in the CTP. 
	\item \textit{Energy Receipt Confirmation (ERC):} This transaction is generated by the smart meter of the consumer when it receives the traded energy. It is assumed that smart meters are tamper resistance and thus ERC cannot be faked. 
\end{itemize}
 Once both CTP and ERC are generated, a smart contract is triggered which pays the price of the energy to the producer. Recall that SPB uses Ethereum as the underlying blockchain and thus employs Ether as the currency for energy price payments.  To reduce the associated overheads and delays, the CTP  is not stored in the blockchain. The miners maintain a separate database, known as CTP database, to store all CTP transactions. To ensure consistency of the CTP database between participating nodes, each miner stores the hash of its CTP database as a field in the block header of newly mined blocks. \par 

The ERC is generated by the smart meter of the energy producer. To protect against malicious nodes that claim to be a smart meter, the participating nodes in the blockchain must be able to verify that a smart meter is genuine. Moreover,  the transactions generated by the smart meter contains privacy-sensitive information of the meter owner, e.g., the energy consumption/production pattern. To address this challenge SPB proposed the notion of a  Certificate of Existence (CoE).   The meter manufacturer populates a key pair in each meter and serves as the CA for all keys. After deployment in the user site, each smart meter creates a number of key pairs and forms a Merkle tree by recursively hashing the  PKs. The root of this tree is then sent to a randomly chosen meter in the network to be signed which serves as the CoE.  The smart meter signs each ERC with the private key corresponding to one of the PKs used to construct the Merkle tree.  The meter also populates the PK and  hashes of Merkle tree leaves to verify existence of the PK in the Merkle tree.  The participating nodes  verify the CoE by: i) verifying the signature and PK of the meter that signed the Merkle tree root, and ii) verifying that the PK exists in the  signed  root hash. 

\section{Proof-of-Concept (PoC)}\label{sec:implementation setup}
\begin{figure*}
	\centerline{\includegraphics[width=16cm,height=10cm,keepaspectratio]{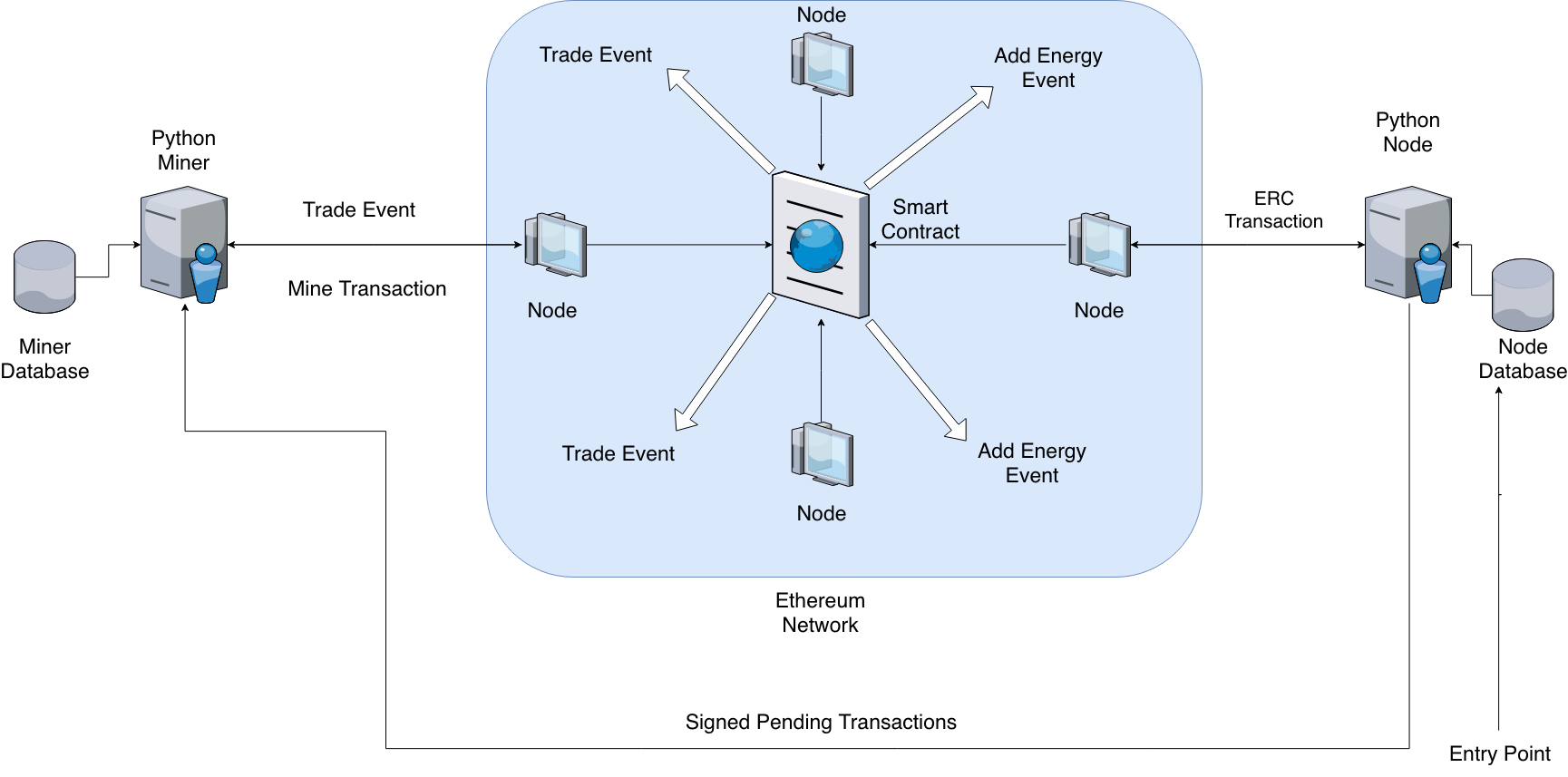}}
	\caption{PoC network architecture.}
	\label{fig:PoCArchitecture}
\end{figure*}

This section outlines the details of PoC implementation of SPB. Figure \ref{fig:PoCArchitecture} depicts the PoC network architecture. Our implementation contains  blockchain network,  a smart contract, and  devices. The blockchain network consists of  the Ethereum testnet which executes the  smart contract and communicates with other devices. All devices (to be discussed later in this section) join the Ethereum private testnet using a Python extension. Recall that in SPB miners maintain a  CTP database to store CTP transactions. To achieve this functionality in the Ethereum testnet,  we implement a Python extension that runs on each node which facilitates propagation, generation and storage of SPB transactions.   \par 
We implement smart contract  using Solidity \cite{Solidity} and run it on Ethereum testnet.  The smart contract contains a function for adding energy to the  balance of each energy producer and verifying the ERC. The energy producer can add energy to his account by sending a transaction to the smart contract with the amount and price of the energy. When a new ERC is generated, the smart contract verifies if the corresponding CTP is present in the CTP database.  In case of a match, the CTP is stored in the blockchain which triggers the payment of the committed funds in the CTP transaction to the producer. Note that, the ERC transaction is not stored in the blockchain. \par 

\begin{figure}
	\centerline{\includegraphics[width=9cm,height=9cm,keepaspectratio]{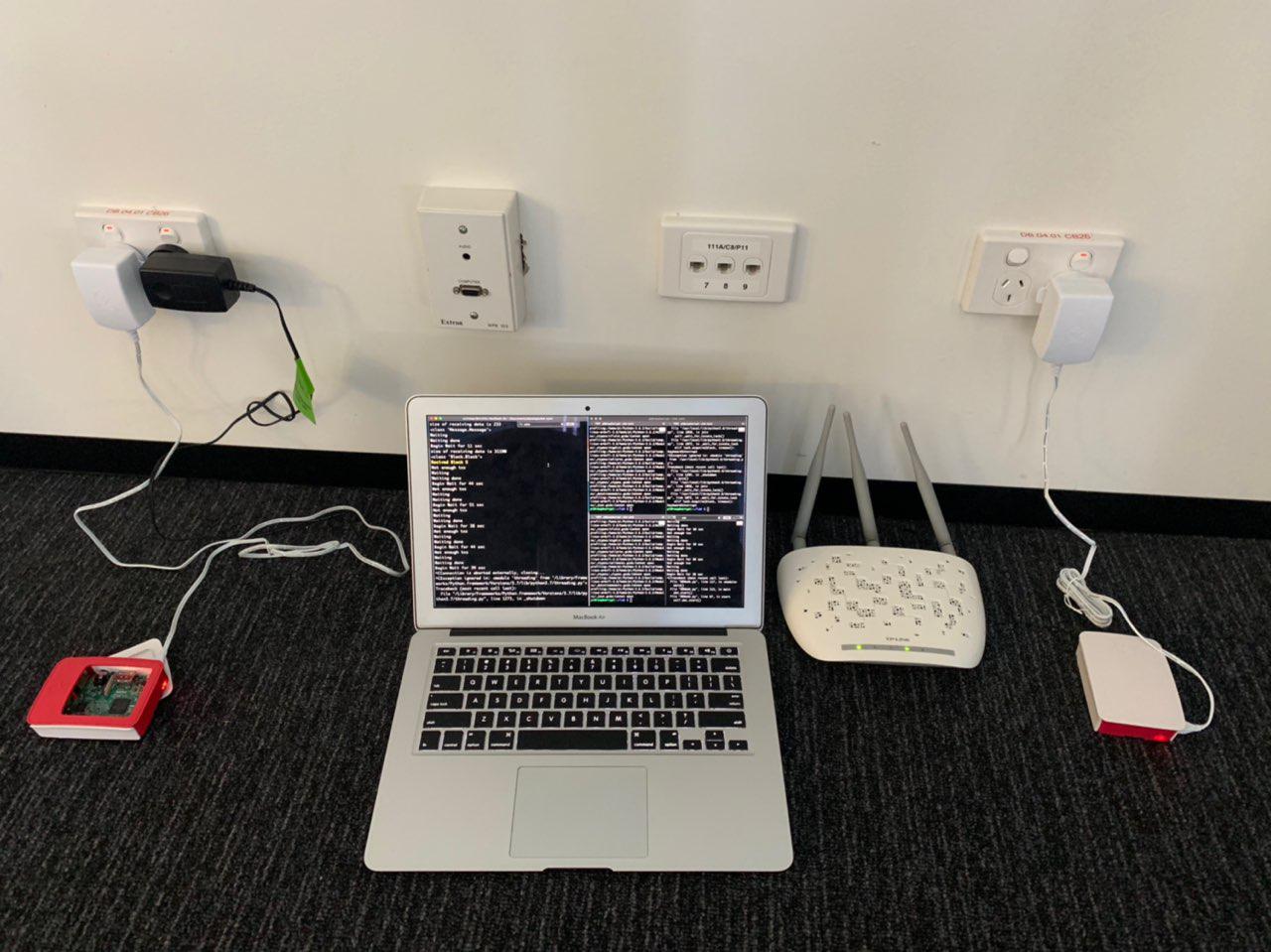}}
	\caption{The implementation setting.}
	\label{fig:implementation-setting}
\end{figure}

We employed two Raspberry Pi3  and a Macbook Pro 2015 with 16 GB RAM and 2.2 GHz Intel Core i7 CPU  as participants in the energy trading as shown in Figure \ref{fig:implementation-setting}. The Pi devices  represent the smart meter of the energy consumer and solar panel of the energy producer while the laptop serves as the miner.  The Pi devices use a Python extension to communicate with Ethereum network and generate transactions. To generate a CTP, we run the following command line in the  Pi device: \par 
\begin{verbatim}
CTP  tx_addr  tx_amount  tx_energy
\end{verbatim}

where \textit{tx\_addr} is the Ethereum address of the producer, and \textit{tx\_amount} is the amount of Ether being committed by the consumer as the energy price. This command creates an entry in the CTP database on the miners through a  Python extension. The Python extension also allocates an  ID to the CTP in the database which is referenced in the corresponding ERC transaction as outlined later in this section. Finally, \textit{tx\_energy} represents the amount of energy to be transfered to the consumer.  The laptop, i.e., the miner, collects transactions and forms new blocks each 15 seconds that is the mining period in Ethereum.   \par 

We run the following command in the Pi device representing the smart meter  of the consumer to generate ERC: 

\begin{verbatim}
ERC CTP_ID  energy_amount
\end{verbatim}
Where \textit{ CTP\_ID} is the ID of the corresponding CTP stored in the CTP database  and \textit{energy\_amount} is the amount of the received energy.  The Python extension then  sends the ERC to the smart contract which  verifies if  the  corresponding CTP transaction exists in the CTP database.  If verified, the  miners add the CTP to the blockchain which in turn triggers the payment of the energy price.     \par 
Each CTP is associated with an expiry\_time which denotes the time period by which the corresponding ERC must be generated. The Python extension code running on the consumer maintains a timer for this duration.   If the producer does not transfer energy before the expiry of the timer, then the consumer uses the Python extension to remove  the  CTP from the CTP database  and thus  releases the corresponding money to the consumer.

\subsection{Applicability}\label{subsec:applicability}

We first study the applicability of SPB for energy trading by implementing two scenarios depicting a reliable and unreliable producer, respectively. The former depicts a properly executed energy trade where the consumer and producer follow the normal steps of operation of SPB. The latter simulates a situation where the producer acts maliciously and does not transfer energy to the consumer after receipt of the CTP.  In both scenarios it is assumed that the energy producer and consumer have reached agreement over the price of the energy, i.e., we exclude the negotiation step.   \par
\begin{figure}
	\centerline{\includegraphics[width=9cm,height=9cm,keepaspectratio]{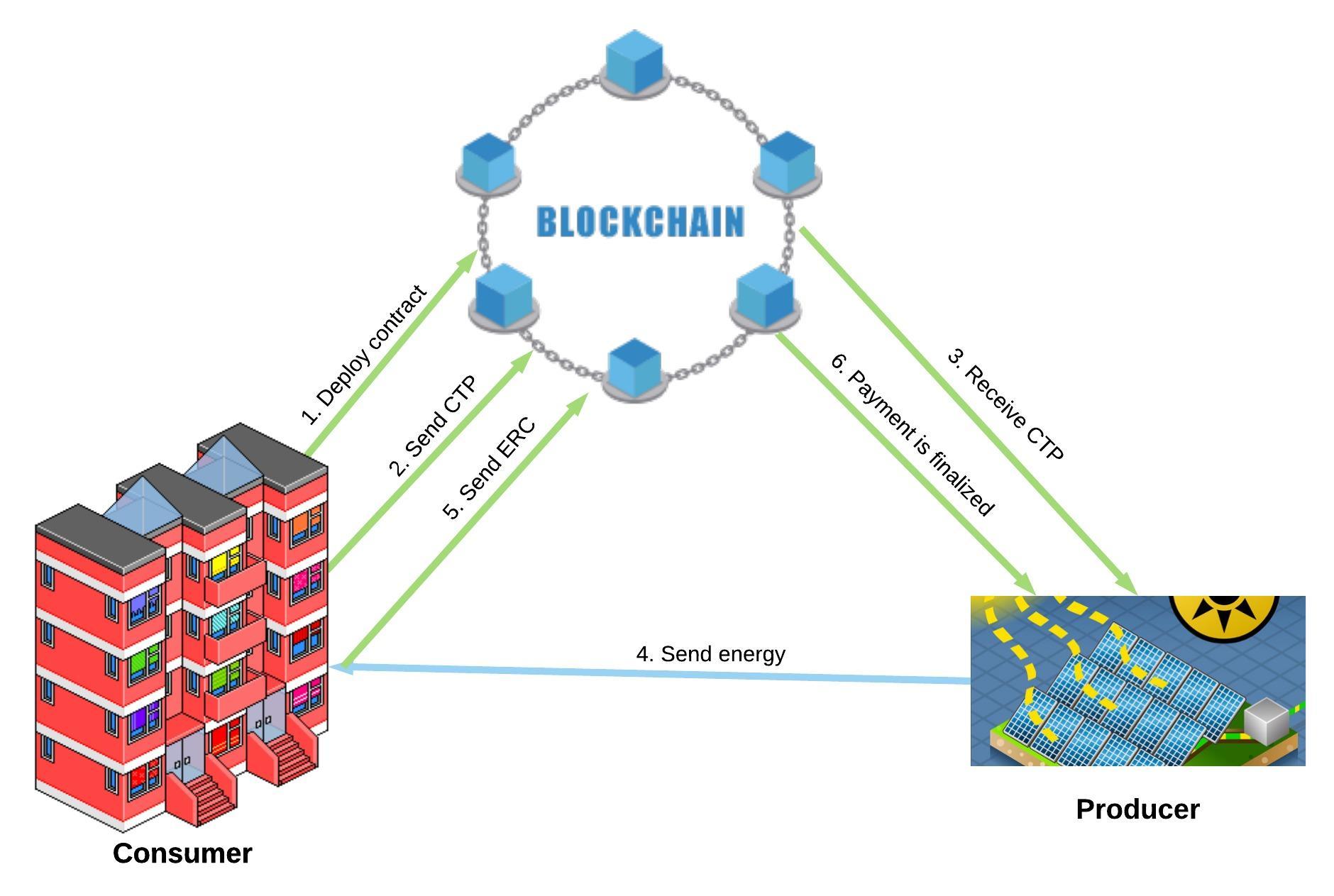}}
	\caption{An overview of steps in reliable energy producer scenario.}
	\label{fig:scenario1}
\end{figure}

 
\textit{Reliable energy producer scenario:} In this scenario the energy consumer and producer fulfill their commitments. Figure \ref{fig:scenario1} shows different steps for energy trading in this scenario. The first step for trading energy is smart contract deployment. The consumer deploys the smart contract in Ethereum testnet (step 1 in Figure  \ref{fig:scenario1}) which is used for all energy trading performed by the consumer. To reduce the associated overhead with storing the smart contract, multiple users may use the same contract as the contract is independent of the contract generator. The energy consumer then generates CTP using the command outlined in Section \ref{sec:implementation setup} (step 2) and broadcasts this transaction to  the network. The Python extension collects the transaction and adds it to the CTP database.  Upon receipt of the CTP transaction (step 3), the energy producer starts transferring energy to the consumer (step 4). After receiving the agreed energy, the smart meter of the consumer generates the ERC (step 5). The ERC completes the second step of the atomic meta-transaction. The smart contract then verifies the ERC. Finally, the negotiated energy price is paid out to the energy producer (step 6). \par 

\begin{figure}
	\centerline{\includegraphics[width=9cm,height=9cm,keepaspectratio]{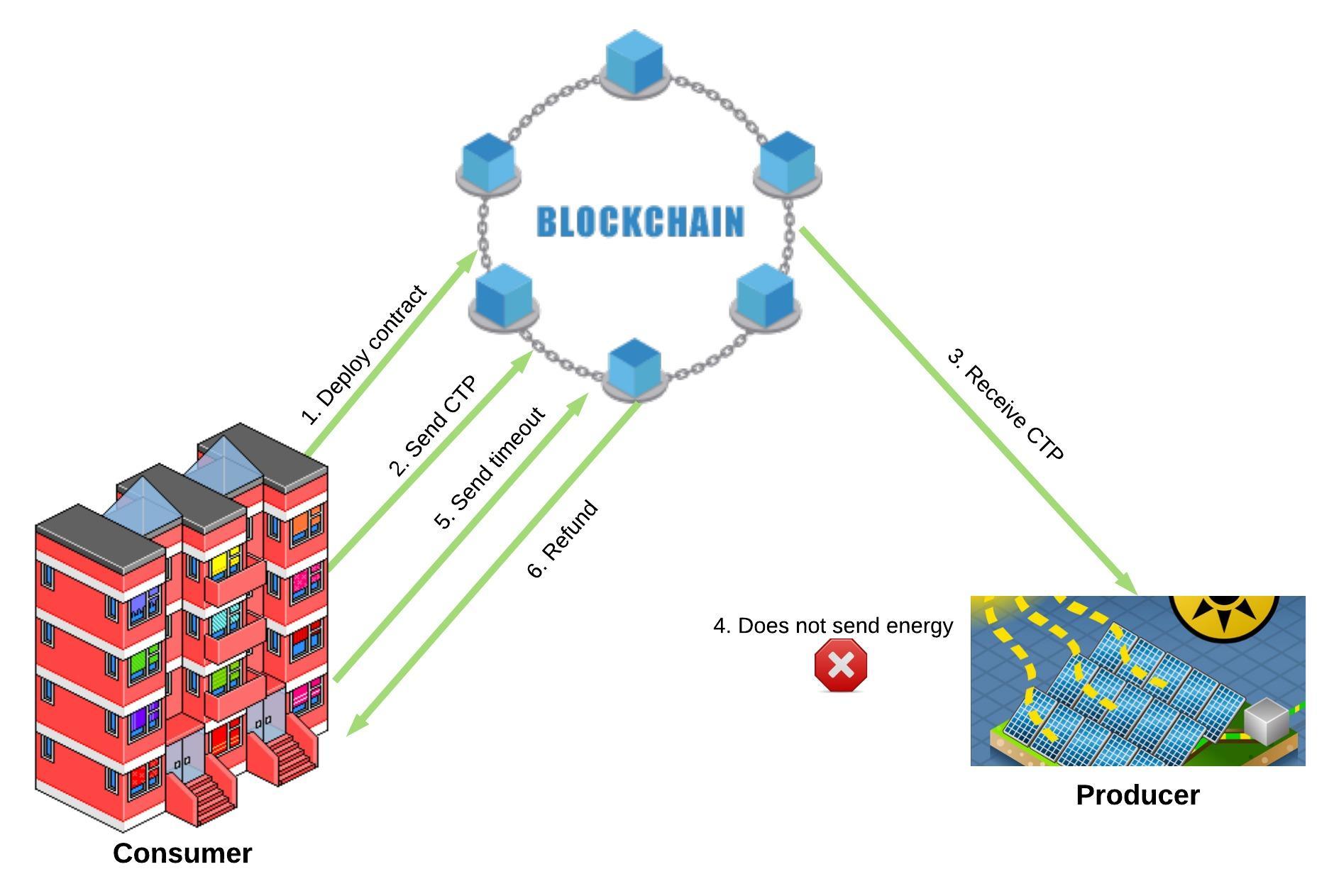}}
	\caption{An overview of steps in unreliable energy producer scenario.}
	\label{fig:scenario2}
\end{figure}

\textit{Unreliable energy producer scenario:} In this scenario the energy consumer does not fulfill  his commitment and  refuses to transfer energy after receiving CTP. Figure \ref{fig:scenario2} shows an overview of this scenario. The first 3 steps are as in reliable energy producer scenario. In step 4, the energy producer does not transfer energy to the consumer. Recall that CTP  does not transfer money to the prosumer account as it requires the corresponding ERC to be generated. The CTP contains an \textit{Expiry time} which is the time period within which the ERC must be generated.  Once the expiry time is reached, the Python extension in the consumer site  sends a timeout request to the miners (step 5).  Finally, the CTP will be removed and the money is refunded to the consumer account (step 6).  \par 

A demonstration of the implementation of these scenarios is available online at \cite{SPBdemonstration}. \par 

\subsection{Performance evaluation} \label{subsec:performanceEvaluation}
In this section, we evaluate the performance of our PoC implementation of SPB and quantify its benefits over a baseline method. The baseline method is similar to conventional energy trading methods using blockchain where the energy trading is conducted by relying solely on a smart contract \cite{DistributedAPP}, discussed in Section \ref{subsec:relatedwork}. The consumer pays the cost of energy to the smart contract as   a TTP. Once the producer transfers energy to the consumer, the consumer confirms receipt of energy and finally  the smart contract pays the energy price to the producer.  We have evaluated the following parameters:\par 
\begin{itemize}
	\item \textit{End-to-End delay:} is the time taken by the entire energy trading process to complete.  As the time for transferring energy is independent of the proposed energy trading framework, we disregard this delay. Thus, the delay basically represents the time taken to generate, broadcast, and  store all transactions necessary to manifest the trade in the blockchain.
		\item \textit{Cost:}  represents the total amount of monetary cost that the end user has to pay as transaction fee for trading energy.  Recall that each transaction involves a transaction fee that is an incentive for the miner to store transactions in the blockchain.
	\item \textit{Throughput:}  represents the time required for a particular number of  energy trading transactions to  be mined. Throughput impacts the delay in mining transactions in the blockchain that is  fundamental metric  particularly in large scale smart grids when the number of energy trading transactions increases.

	\item \textit{Blockchain size: }  represents the memory footprint of the blockchain for storing all energy trading related transactions. This particularly impacts the blockchain management cost and overhead for the participating nodes and affects scalability.
\end{itemize}

For our evaluations, we use the network setup  shown in Figure \ref{fig:implementation-setting}. The results presented in the rest of the paper are averaged over 100 runs of the experiments.   In the rest of this section we discuss the PoC results. \par 
\subsubsection{End-to-End Delay}
\begin{figure}
	\centerline{\includegraphics[width=7cm,height=7cm,keepaspectratio]{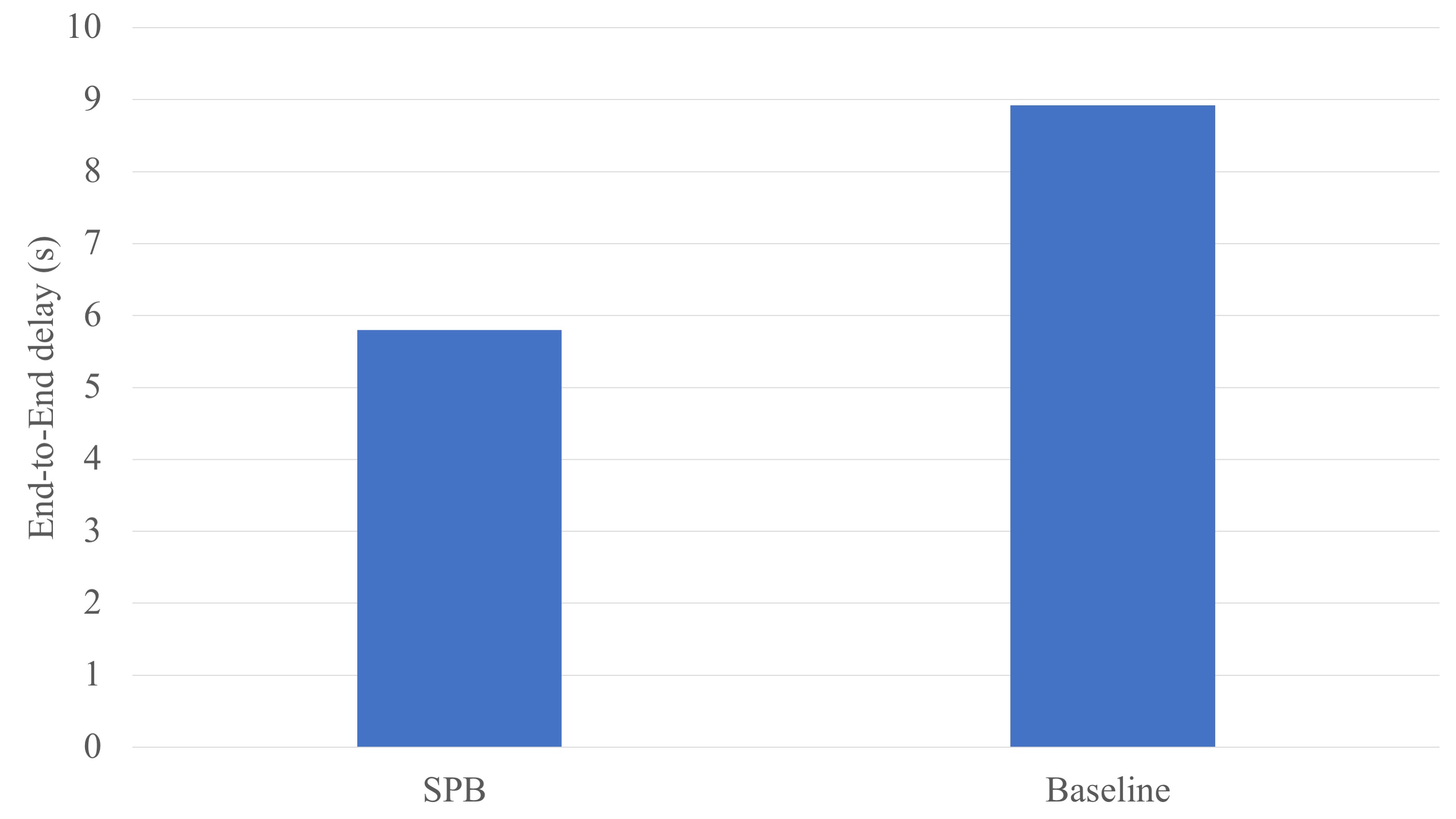}}
	\caption{Evaluation of end-to-end delay.}
	\label{fig:delay}
\end{figure}

 Figure \ref{fig:delay} depicts the delay for SPB and the baseline.  SPB reduces delay for energy trading  as only one transaction is stored in the blockchain compared to three transactions in the baseline.  Recall that once the user generates the CTP, it is stored in a separate database rather than in the blockchain, which avoids the need for its inclusion in the consensus process and in turn reduces delay. Figure \ref{fig:delay} shows SPB reduces the end-to-end delay by 35\% compared to baseline.  \par 

\subsubsection{Cost}
\begin{figure}
	\centerline{\includegraphics[width=9cm,height=9cm,keepaspectratio]{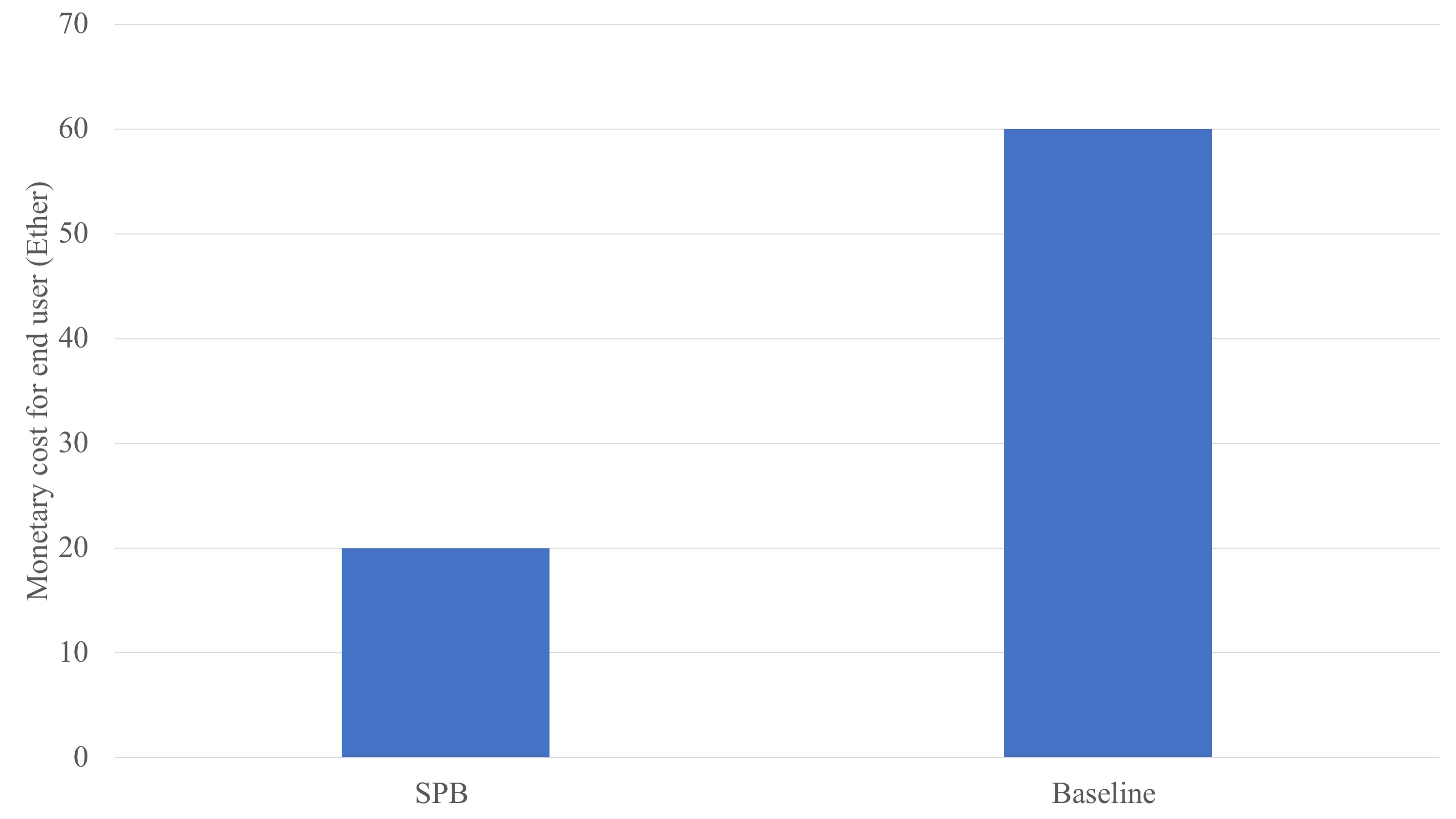}}
	\caption{Evaluation of monetary cost for the end user.}
	\label{fig:cost}
\end{figure}

 In the baseline method three transactions must be   stored in the blockchain for each energy trade which include:  the smart contract, the payment made by the consumer to the smart contract, and the smart contract payment to the producer. However, in SPB only one transaction needs to be stored which is the final CTP once the smart meter of the consumer generates the ERC.  It is assumed that the cost for storing each transaction is 20 Ether.  Figure \ref{fig:cost} illustrates the monetary cost incurred by both methods. As can be seen SPB  reduces the monetary cost for the end user to 20 Ethers as compared to baseline with 60 Ethers.  \par 

\subsubsection{Throughput}
\begin{figure}
	\centerline{\includegraphics[width=7cm,height=7cm,keepaspectratio]{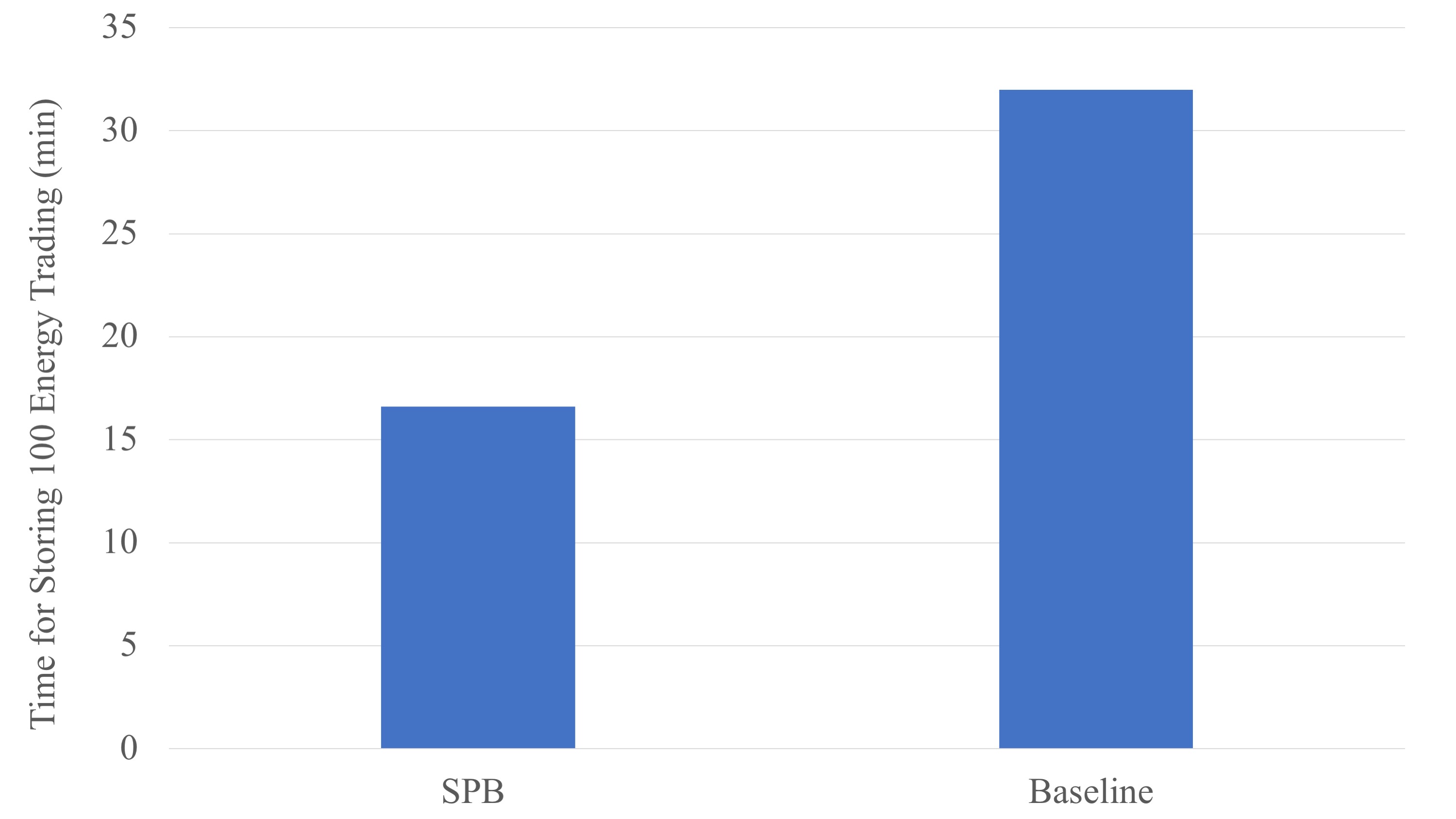}}
	\caption{Evaluation of throughput.}
	\label{fig:throughput}
\end{figure}

Figure \ref{fig:throughput} represents the network throughput for SPB and the baseline. Evidently, in SPB it takes 16.6 minutes for 100 energy trading transactions to be stored in the blockchain while in the baseline it takes 32 minutes to store the same number of energy trading transactions.  Recall that in SPB only one transaction is stored in the blockchain compared to three transactions in baseline for each energy trading. SPB increases the number of energy trading  transactions that can be stored in the blockchain in the same time period by 48\% compared to the baseline, which in turn decreases delay in energy trading particularly in large scale smart grid.

\subsubsection{Blockchain size}
\begin{figure}
	\centerline{\includegraphics[width=9cm,height=9cm,keepaspectratio]{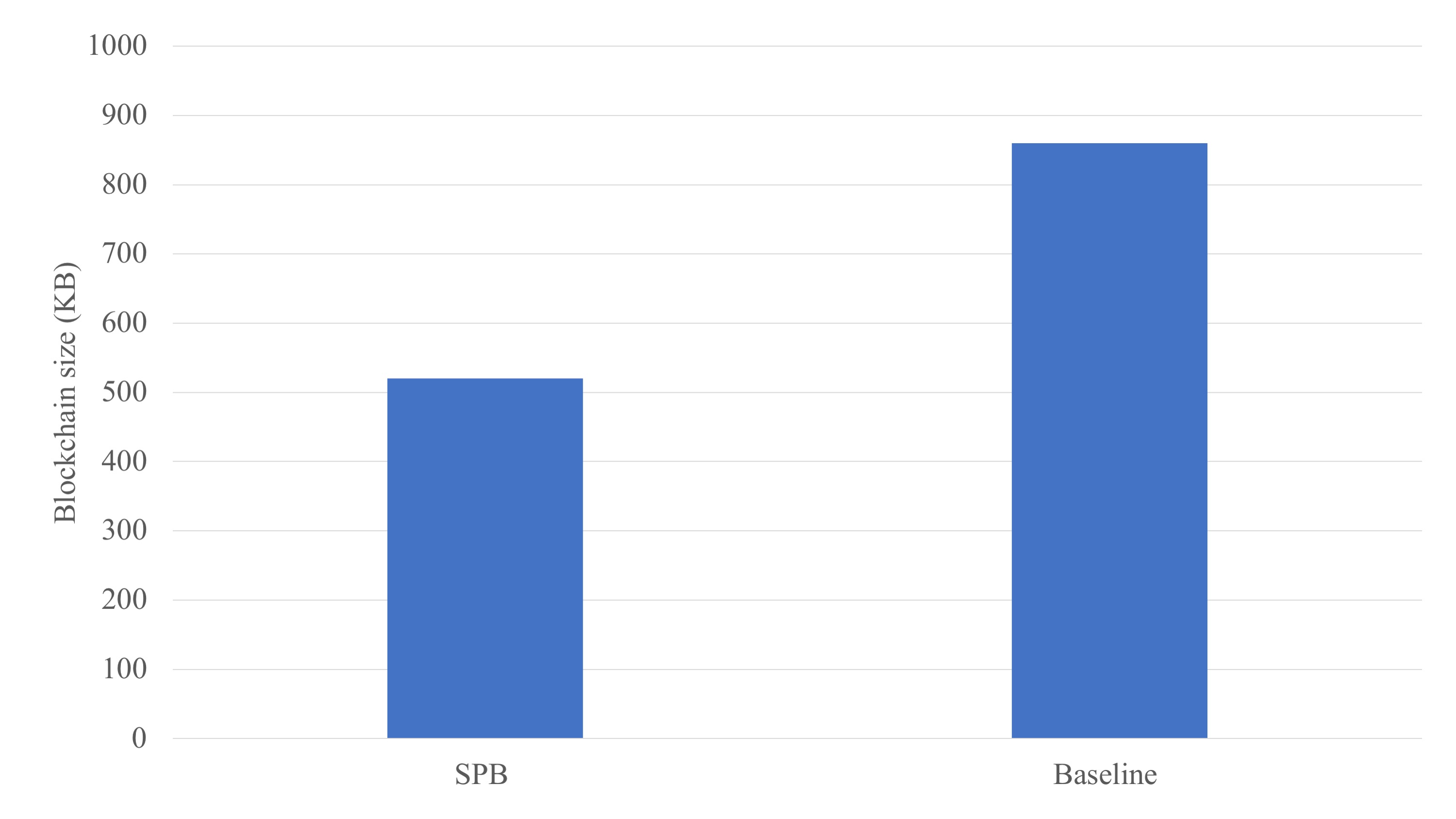}}
	\caption{Evaluation of blockchain size.}
	\label{fig:bcsize}
\end{figure}

SPB stores fewer transactions as compared to the baseline in the blockchain.   Figure \ref{fig:bcsize}  illustrates the memory footprint for the two methods.  The blockchain size is measured once 100 energy trades have been actioned in both methods. As shown, the blockchain size reaches 520 KB in SPB as compared to 860 KB in the baseline. In the baseline for each energy trading three transactions are stored, while in SPB only one transaction, that is the CTP, is stored in the blockchain. SPB reduces  the size of the blockchain which potentially reduces blockchain management cost incurred in the miners  and increases scalability. \par 

\section{Conclusion and Future Work}\label{sec:conclusion}
Recently blockchain applications in smart energy trading have received tremendous attention due to its salient features that include decentralization, security, and  privacy. However, existing blockchain-based solutions suffer from lack of privacy, reliance on Trusted Third Parties (TTP), and blockchain overheads. In this paper,  we presented a Proof of Concept (PoC) implementation of a Secure Private Blockchain-based (SPB) energy trading framework. We implemented SPB on Ethereum private network. We built a Python network that enables us to modify Ethereum network behavior to be consistent with SPB. The PoC results showed SPB  reduces cost, blockchain size, and processing time for energy trading. \par 
In our future work, we plan to extend the current research in the  following directions: (1) develop a new blockchain-based and energy-oriented virtual currency system that   integrates a new consensus algorithm  based on the prosumer's demand response effort;  (2) develop a peer-to-peer energy trading market structure, which is automated by both SPB and building energy management systems; (3)  evaluate the scalability of SPB using a larger testbed where multiple energy consumers and producers negotiate and trade energy.

\bibliographystyle{IEEEtran}
\bibliography{bare_jrnl_compsoc}
\end{document}